\documentclass[aps,pra,twocolumn,groupedaddress,showpacs,notitlepage]{revtex4-1}
\usepackage{hyperref}
\usepackage{amsmath}
\usepackage{graphicx}

\usepackage[T1]{fontenc}
\usepackage[latin1]{inputenc}

\newcommand{\be}{\begin{equation}}
\newcommand{\ee}{\end{equation}}

\begin{document}

\title{Lattice topology and spontaneous parametric down-conversion in quadratic nonlinear waveguide arrays}

\author{Daniel Leykam}

\author{Alexander S. Solntsev}
\author{Andrey A. Sukhorukov}
\author{Anton S. Desyatnikov}

\affiliation{Nonlinear Physics Centre, Research School of Physics and Engineering, The Australian National University, Canberra ACT 0200, Australia
}

\date{\today}

\begin{abstract}
We analyze spontaneous parametric down-conversion in various experimentally feasible 1D quadratic nonlinear waveguide arrays, with emphasis on the relationship between the lattice's topological invariants and the biphoton correlations. Nontrivial topology results in a nontrivial ``winding'' of the array's Bloch waves, which introduces additional selection rules for the generation of biphotons. These selection rules are in addition to, and independent of existing control using the pump beam's spatial profile and phase matching conditions. In finite lattices, nontrivial topology produces single photon edge modes, resulting in ``hybrid'' biphoton edge modes, with one photon localized at the edge and the other propagating into the bulk. When the single photon band gap is sufficiently large, these hybrid biphoton modes reside in a band gap of the bulk biphoton Bloch wave spectrum. Numerical simulations support our analytical results.
\end{abstract}

\pacs{42.65.Lm, 42.50.Dv, 42.65.Wi}

\maketitle

\section{Introduction}

Spontaneous parametric down-conversion (SPDC) is an important way to generate pairs of photons exhibiting quantum correlations, with applications ranging from fundamental tests of quantum theory (Bell tests) to quantum cryptography and information processing~\cite{grice1997,bell_test,QIP}. Genuinely quantum behaviour and scalability require a high fidelity of the photon pairs, which is limited if they are generated and shaped by bulk optical components. Hence there is currently strong interest in implementing SPDC in integrated optical devices~\cite{integrated_1,integrated_2,integrated_3}.

SPDC in nonlinear waveguide arrays has been proposed as a tool to tailor biphoton quantum correlations in integrated optics~\cite{SPDC_arrays,driven_walks,SPDC_array_2,SPDC_array_3,grafe2012,markin2013}. It offers many advantages over bulk components; biphotons are generated directly in the device, so there are no input coupling losses. The biphoton spectrum and correlations can also be readily controlled via the pump beam's spatial profile and the diffraction or quantum walk of the generated photon pairs through the array~\cite{biphoton_lattice,peruzzo2010,poulios2014}. This concept was recently demonstrated in experiments in lithium niobate waveguide arrays~\cite{iwanow,kruse2013,solntsev2014,SPDC_array_1}.

So far however, SPDC was only studied in homogeneous waveguide arrays, or arrays with a single defect~\cite{kruse2013,SPDC_arrays,driven_walks}. In both cases only a single Bloch band is relevant. It is interesting therefore to explore the opportunities for controlling SPDC and biphoton correlations offered by \emph{modulated} arrays with \emph{multiple} Bloch bands~\cite{modulated_arrays}.

Given the endless possibilities in designing modulated waveguide arrays, it is useful to group them into different classes sharing similar properties. One way to do this is using \emph{topological invariants}, which has lead to a wide range of breakthroughs both in the fundamental band theory of solids, to devices using ``topologically protected'' surface states that are robust against disorder~\cite{topological_review,topological_review_1,topological_review_2}. These ideas are now attracting interest in optics, and several photonic analogues of these topological condensed matter systems were recently demonstrated in experiments~\cite{topological_photonics}.

Here we explore how \emph{lattice topology} can provide an additional tool to control correlations of biphotons generated in quadratic nonlinear waveguide arrays. The basic idea is that nontrivial topology is associated with a nontrivial ``winding'' of the lattice's Bloch waves, and this winding can lead to selection rules controlling which biphoton modes are strongly excited. In principle, the mode winding can be completely independent from the array's dispersion relation (phase matching conditions), so it offers another degree of freedom to control biphoton correlations in integrated optics. This topology is inherently robust against fabrication disorder. In some cases lattices with nontrivial topology also host protected edge modes, which produce bands of ``hybrid'' biphoton states exhibiting entanglement between localized and propagating modes. We demonstrate the feasibility of these ideas by carrying out numerical simulations of various one dimensional (1D) topological lattices under the tight binding approximation. The model and parameter regimes are accessible in current state of the art experiments.

In Sec.~\ref{sec:general_theory} we review the theory of SPDC in waveguide arrays, generalising to multi-band (modulated) arrays. In Sec.~\ref{sec:2_band} we make some general statements on the role of topology in two band, 1D models. Following this, as a concrete example we consider in detail the Su-Schrieffer-Heeger (SSH) model in Sec.~\ref{sec:SSH_model}, which illustrates main features and is a practical, experimentally realisable example. Sec.~\ref{sec:modulated_array} compares these results against a binary lattice where the waveguide depths are modulated, which is an example of a ``nontopological'' model because it lacks the required symmetry. We conclude in Sec.~\ref{sec:conclusion} with a summary and discussion of future directions.

\section{Setup and observables}
\label{sec:general_theory}

\begin{figure}

\includegraphics[width=\columnwidth]{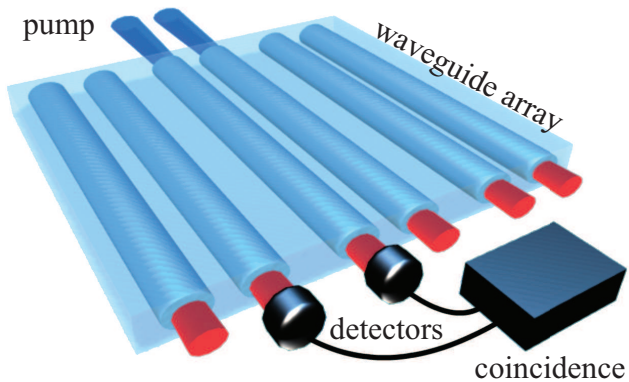}

\caption{(Color online) Schematic of quadratic nonlinear waveguide array. Pump beam generates photon pairs that diffract to neighboring waveguides. Quantum correlations of the photon pairs can be measured via coincidences of two single photon detectors.}

\label{fig:schematic}

\end{figure}

We consider the process schematically illustrated in Fig.~\ref{fig:schematic}. A pump beam at frequency $\omega_p$ propagates through a quadratic nonlinear waveguide array. Nonlinear wave mixing combined with quantum fluctuations can convert a pump photon into two lower frequency photons called the signal and idler. The state of these photons is described by a biphoton wavefunction which evolves as they propagate through the array. The quantum correlations of the signal and idler photons leaving the array can be observed through coincidence measurements of a pair of single photon detectors.

Theoretically, we employ the formalism of Refs.~\cite{christ2009,giuseppe2002,SPDC_arrays}, considering type-1 near-degenerate SPDC under continuous wave pumping at frequency $\omega_p$, such that phase matching occurs when $\omega_p \approx 2 \omega_{s,i}$, where $\omega_{s} \approx \omega_i$ are the signal and idler photon frequencies.  This can be implemented in experiments by placing an appropriately chosen spectral filter at the array output.

Close to degeneracy, the waveguide coupling coefficients for the signal and idler photons are approximately the same, ie. $C_{s,i} \equiv C$. On the other hand, the higher frequency pump beam experiences much stronger confinement and hence weaker coupling $C_p$ between neighbouring waveguides. A good approximation for recent experiments is $C_p \approx 0$, such that coupling of pump photons between waveguides can be neglected~\cite{solntsev2014}. Consequently, under the undepleted pump approximation and choosing a frame rotating at the pump frequency, the pump beam profile remains constant along the waveguide array, simplifying the theoretical analysis considerably. 

Under these conditions, the evolution of the biphoton wavefunction can be described by a Hamiltonian $\hat{H} = \hat{H}^{(\mathrm{QW})} + \hat{H}^{(\mathrm{SPDC})}$. $\hat{H}^{(\mathrm{QW})}$ accounts for the linear diffraction (quantum walk) of biphotons through the waveguide array, and $\hat{H}^{(\mathrm{SPDC})}$ is a gain term accounting for their generation via SPDC. In normalized units with $\hbar = 1$,
\begin{align}
\hat{H}^{(\mathrm{QW})} &= \sum_{n, m} \left[ \hat{a}_{m}^{(s)\dagger} H_{mn}  \hat{a}_{n}^{(s)} + \hat{a}_{m}^{(i)\dagger} H_{mn} \hat{a}_{n}^{(i)}\right], \label{H_SPDC-1} \\
\hat{H}^{(\mathrm{SPDC})} &= i \gamma \sum_{n_p} E_{n_p}^{(p)} \hat{a}_{n_p}^{(s)\dagger} \hat{a}_{n_p}^{(i)\dagger} + \mathrm{H.c.}, \label{H_SPDC-2}
\end{align}
where $\hat{a}_n^{(s,i)\dagger} (\hat{a}_n^{(s,i)})$ creates (destroys) a signal or idler photon at the $n$th waveguide, $\gamma$ is the nonlinear coefficient, $E_{n_p}^{(p)}$ is the pump amplitude in waveguide $n_p$, and $H_{nm}$ are elements of the waveguide array's tight binding Hamiltonian. Diagonal elements $H_{nn}$ account for the propagation constant of the signal/idler photons in the $n$th waveguide; off-diagonal elements describe evanescent coupling between waveguides.

We assume there is no decoherence or loss, such that in the absence of multiple photon pairs being generated simultaneously, the biphoton state $\mid \Psi \rangle$ is pure and evolves according to the Schr\"odinger equation~\cite{SPDC_array_3},
\be 
i \partial_z \mid \Psi \rangle \simeq \left[ \hat{H}^{(\mathrm{QW})} + \hat{H}^{(\mathrm{SPDC})} \right] \left( \mid \Psi \rangle + \mid 0 \rangle \right), \label{schrodinger-equation}
\ee
where $\mid 0 \rangle$ is the vacuum state. We will solve this equation and reveal the effect of band structure topology by transforming to the eigenbasis of $\hat{H}^{(\mathrm{QW})}$, ie. the lattice's Bloch wave basis.

Consider a waveguide superlattice with a unit cell consisting of $N$ waveguides. It is convenient to introduce the vector notation $\hat{\mathbf{a}}_{n} = (\hat{a}_{n,1},...,\hat{a}_{n,N})$, where now $n$ numbers the unit cell and $\hat{a}_{n,m}$ is the annihilation operator for the $m$th sublattice. Eq.~\eqref{H_SPDC-1} is recast as
\be 
\hat{H}^{(\mathrm{QW})} = \sum_{n, m} \left[ \hat{\mathbf{a}}_{m}^{(s)\dagger} \hat{H}_{mn}  \hat{\mathbf{a}}_{n}^{(s)} + \hat{\mathbf{a}}_{m}^{(i)\dagger} \hat{H}_{mn}  \hat{\mathbf{a}}_{n}^{(i)}\right], \label{H_SPDC-1b}
\ee
where now each $\hat{H}_{mn}$ is promoted to an $N\times N$ matrix, with off-diagonal elements accounting for coupling between the different sublattices. $\hat{H}_{mn} = \hat{H}_{m+1n}=\hat{H}_{mn+1}$ is periodic, such that transforming to reciprocal space $\hat{\mathbf{a}} ( k ) = \sum_n \hat{\mathbf{a}}_n e^{i k n}$ puts $\hat{H}^{(\mathrm{QW})}$ into block diagonal form,
\be 
\hat{H}^{(\mathrm{QW})} (k) = \sum_n \hat{H}_{0n} e^{i k n}. \label{Bloch_Hamiltonian}
\ee
The eigenvectors of $\hat{H}^{(\mathrm{QW})}(k)$ are the superlattice's Bloch functions $\mathbf{u}_p ( k )$; eigenmodes of $\hat{H}^{(\mathrm{QW})}$ are Bloch waves, constructed by
\be 
\hat{B}_p ( k ) = \sum_n \hat{\mathbf{a}}_n \cdot \mathbf{u}_p ( k ) e^{i k n}, \label{bloch_mode}
\ee
where $\cdot$ denotes the usual dot product, and $p=1,...,N$ is the band index. In this Bloch wave basis, $\hat{H}^{(\mathrm{QW})}$ takes the simple diagonal form
\begin{align}
\hat{H}^{(\mathrm{QW})} &= \sum_{p_s} \int_{-\pi}^{\pi} dk_s \beta_{p_s} (k_s )\hat{B}_{p_s}^{\dagger} (k_s ) \hat{B}_{p_s} (k_s ) \nonumber \\ &+ \sum_{p_i} \int_{-\pi}^{\pi} dk_i \beta_{p_i} (k_i )\hat{B}_{p_i}^{\dagger} (k_i ) \hat{B}_{p_i} (k_i ), \label{QW_bloch_basis} 
\end{align}
where $\beta_p ( k )$ is the propagation constant of the Bloch wave in band $p$ with crystal momentum $k$. To obtain $\hat{H}^{(\mathrm{SPDC})}$ in this basis, we invert Eq.~\eqref{bloch_mode},
\be 
\hat{\mathbf{a}}_n = \sum_p \int_{-\pi}^{\pi} dk \mathbf{u}_p^* ( k ) \hat{B}_p ( k ) e^{-i k n},
\ee
and substitute into Eq.~\eqref{H_SPDC-2}. Writing the pump amplitude in vector form in terms of its sublattice components, $\mathbf{E}_{n}^{(p)} = (E_{n,1},...,E_{n,N})$, and applying a Fourier transform, $\mathbf{E}^{(p)}(k_p ) = \sum_{n} \mathbf{E}_{n}^{(p)} e^{i k_p n}$, we obtain
\be 
\hat{H}^{(\mathrm{SPDC})} = i \sum_{p_s,p_i} \int dk_s dk_i \Gamma_{p_s,p_i} (k_s,k_i) \hat{B}_{p_s}^{\dagger} ( k_s )\hat{B}_{p_i}^{\dagger} ( k_i ), \label{spdc_bloch_basis}
\ee
where
\be 
\Gamma_{p_s,p_i} (k_s,k_i) = \gamma \sum_{j=1}^N E_j^{(p)} (k_s + k_i) u_{p_s,j} (k_s ) u_{p_i,j} (k_i ), \label{coupling_efficiency}
\ee
is the \emph{coupling efficiency} into the biphoton Bloch wave. The summation is over the $N$ sublattices forming the superlattice. Eq.~\eqref{spdc_bloch_basis} is also diagonal in the Bloch wave basis, thus in a similar manner to Ref.~\cite{SPDC_arrays} we can integrate Eq.~\eqref{schrodinger-equation} to obtain the output biphoton wavefunction (up to an overall normalization factor),
\begin{align} 
\mid \Psi \rangle =& \sum_{p_s,p_i} \int dk_s dk_i \Gamma_{p_s,p_i} (k_s, k_i ) L\mathrm{sinc} ( \Delta \beta L / 2 ) \nonumber \\
&\times \exp ( - i \Delta \beta L / 2 ) \hat{B}_{p_s}^{\dagger} ( k_s ) 
\hat{B}_{p_i}^{\dagger} ( k_i ) \mid 0, 0 \rangle, \label{schrodinger_solution}
\end{align}
where $\Delta \beta = \Delta \beta^{(0)} - \beta_{p_s} (k_s) - \beta_{p_i} (k_i)$ is the phase mismatch into the biphoton Bloch wave, $\Delta \beta^{(0)}$ is the single waveguide phase mismatch, and $L$ is the propagation length.

Eq.~\eqref{schrodinger_solution} tells us that two factors determine whether a biphoton Bloch wave is strongly populated: how close the mode is to phase matching (small $|\Delta \beta|$), and how strongly the pump beam profile is matched to the mode's transverse profile (large $|\Gamma|$). Let us now discuss how the lattice topology can affect each of these.

The phase matching condition $\Delta \beta  = 0$ depends only on the biphoton mode eigenvalues. Since the lattice topology is completely independent of the spatial dispersion $\beta ( k )$, in an infinite lattice the phase matching is insensitive to the topology: it cannot distinguish between two topologically distinct lattices~\cite{topological_review}. On the other hand, in a \emph{finite} lattice, nontrivial topology can result in ``topologically protected'', exponentially localized edge modes~\cite{topological_review}. The phase matching condition is sensitive to these edge modes.

The coupling efficiency $|\Gamma|$ clearly depends on the Bloch function profiles $\mathbf{u}_p ( k )$ via Eq.~\eqref{coupling_efficiency}. Thus, we expect nontrivial ``winding'' or topology of the Bloch functions to have some effect, even in an infinite lattice.

While this Bloch wave decomposition is a convenient way to theoretically study SPDC in a superlattice, unfortunately $\Gamma$ and the Bloch functions $\mathbf{u}_p ( k )$ are not directly observable in experiments. Instead, what is typically measured is the magnitude of the biphoton wavefunction Eq.~\eqref{schrodinger_solution}, in either real or Fourier space, using coincidence measurements from a pair of single photon detectors. Therefore, instead of the Bloch wave basis indexed by band number $p$ and crystal momentum $k$, we also need to consider the output in real and momentum space. For the latter, we use an extended Brillouin zone represenation, allowing $k \in [-N\pi,N\pi]$ to lie in the first $N$ Brillouin zones, and using the Fourier amplitudes in the $p$th Brillouin zone as a proxy for the Bloch wave amplitude in the $p$th band~\cite{kittel,yariv}.

We would also like to quantify how ``quantum'' a given biphoton state is, and whether the lattice topology influences the entanglement of the generated photons. One useful measurement of quantumness is the Schmidt number~\cite{schmidt_number}, obtained via the singular value decomposition of $\mid \Psi \rangle$,
\be 
\mid \Psi \rangle = \sum_j \sqrt{\lambda_j} \mid v^{(s)}_j \rangle \otimes \mid v^{(i)}_j \rangle,
\ee
where $\mid v_j \rangle$ are the Schmidt modes, $\lambda_j > 0$ are normalised such that $\sum_j \lambda_j = 1$, and we define the Schmidt number as $K = \sum_j 1/\lambda_j^2$, which measures the number of entangled modes.

\section{Two band models}
\label{sec:2_band}

The simplest case allowing for nontrivial topology is $N=2$ band models, which often form a good approximation to more complicated systems. The most general two band Bloch Hamiltonian Eq.~\eqref{Bloch_Hamiltonian} is~\cite{topological_review}
\be 
\hat{H}^{(\mathrm{QW})} ( k ) = \left( \begin{array}{cc} d_z & d_x - i d_y \\ d_x + i d_y & -d_z \end{array} \right) = \mathbf{d}(k) \cdot \mathbf{\hat{\sigma } }, ~\label{2_band_hamiltonian}
\ee
where $\mathbf{\hat{\sigma}} = (\sigma_x, \sigma_y, \sigma_z )$ is a vector consisting of the three Pauli matrices, and we will see in the following that the vector $\mathbf{d}(k) = (d_x ( k),d_y ( k ),d_z ( k ))$ provides a convenient way to visualise both the Bloch functions and their topology. Diagonal elements of the Bloch Hamiltonian $\hat{H}^{(\mathrm{QW})} ( k )$ account for coupling between waveguides belonging to the same sublattice and their propagation constants, while off-diagonal elements account for coupling between different sublattices.

Diagonalizing Eq.~\eqref{2_band_hamiltonian}, we obtain the Bloch wave eigenvalues and corresponding Bloch functions,
\begin{align}
\beta_{\pm} ( k ) &= \pm |\mathbf{d}(k) |, \label{general_prop_constant} \\
\mathbf{u}_{\pm} (k ) &= \left( e^{-i \phi (k)} \sin \frac{\varphi(k)}{2} , \pm \cos \frac{\varphi(k)}{2} \right), \label{general_modes}
\end{align}
where we have introduced the spherical polar angles $\varphi ( k )$ and $\phi ( k )$ corresponding to the direction $\hat{\mathbf{d}}(k) \equiv \mathbf{d}/|\mathbf{d}| = ( \cos \phi \sin \varphi, \sin \phi \sin \varphi, \cos \varphi) $. We assume the two bands are separated by a gap, ie. $\beta_{\pm} \ne 0$, such that $|\mathbf{d}(k) |\ne 0$ for all $k$ and the angles $\phi, \varphi$ are always well-defined.

\begin{figure}

\includegraphics[width=\columnwidth]{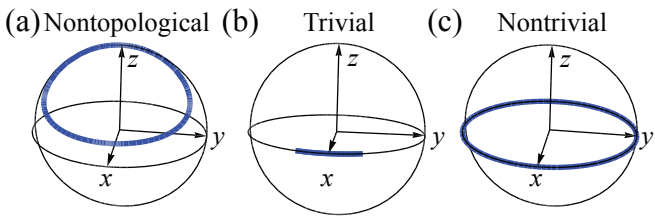}

\caption{(Color online) Bloch sphere representation of Bloch function topology. Bloch functions $\mathbf{u}_p ( k )$ form a closed loop on Bloch sphere surface as $k$ traverses the 1D Brillouin zone, eg. (a). (b,c) Topological phases phases are well-defined when $\mathbf{u}_p ( k )$ is constrained by symmetry to lie on a circle. Here chiral symmetry restricts $\mathbf{u}_p ( k )$ to the equatorial plane. The curve completely winds around the circle in the nontrivial phase (c), and it cannot be continuously deformed to (b).}

\label{bloch_sphere}

\end{figure}

The Bloch sphere provides a simple way to visualize the Bloch functions and their topology. The Bloch functions are mapped to points on the sphere's surface specified by the pair of angles $(\phi, \varphi)$. Since the Bloch Hamiltonian is periodic, $\hat{H}^{(\mathrm{QW})}(k) = \hat{H}^{(\mathrm{QW})}(k + 2 \pi)$, $\phi(k)$ and $\varphi(k)$ are periodic as well. Therefore as $k$ traverses the Brillouin zone, $(\phi, \varphi)$ maps out a closed curve on the sphere surface, see eg. Fig.~\ref{bloch_sphere}(a). This allows the topology of the Bloch functions to be mapped to the topology of curves on a sphere.

The topology is trivial if the closed curve can be continuously deformed to a point. In a general two-band model without any symmetries, such that $d_{x,y,z}$ are unconstrained and $(\phi,\varphi)$ can take any value, this is always possible and the topology is trivial. Hence nontrivial topology in a 1D lattice requires some symmetry. An example is the ``chiral'' symmetry $d_z ( k ) = 0$~\cite{topological_review}, which restricts $(\phi,\varphi)$ to the equatorial plane in Fig.~\ref{bloch_sphere}(b,c). As long as the two Bloch bands are separated by a gap, $\beta_{\pm} \ne 0$, there is no way to continuously deform the trivial curve (b) such that it has the nontrivial winding around the equator in (c). 

In a finite lattice, this nontrivial winding results in protected edge modes with propagation constant $\beta = 0$ in the middle of the gap between the two bands. Their transverse profiles decay exponentially away from the edge of the lattice, at a rate determined by the size of the gap~\cite{delplace2011}. The modes are protected in the sense that they cannot be destroyed by any perturbation that respects the chiral symmetry $d_z = 0$ as long as the two bands remain separated by a gap~\cite{topological_review}. The topological invariant associated with this protection is the Zak phase~\cite{zak_phase}.

Turning to properties of biphoton modes, the signal and idler photons can excite different combinations of the two single photon bands with propagation constants given by Eq.~\eqref{general_prop_constant}. Thus there are four biphoton bands with energies $\pm (|\mathbf{d}(k_s)| + |\mathbf{d}(k_i)|)$ and $\pm (|\mathbf{d}(k_s)| - |\mathbf{d}(k_i)|)$, corresponding to the signal and idler photons exciting the same and different single photon bands respectively. If signal and idler photons are near-degenerate, such that $\hat{H}_s \approx \hat{H}_i$, then the $|\mathbf{d}(k_s)| - |\mathbf{d}(k_i)|$ and $-|\mathbf{d}(k_s)| + |\mathbf{d}(k_i)|$ bands overlap, centred at $\beta = 0$. When the width of a single photon band is larger than the band gap, the biphoton spectrum is gapless. To show this, we note that $\Delta = 2 \mathrm{min}(|\mathbf{d}(k_s)|)$ is the size of the single photon band gap and let $w = \mathrm{max}(|\mathbf{d}(k_s)|) - \mathrm{min}(|\mathbf{d}(k_s)|)$ be the band width. Then the bottom of the top band, $\Delta$, is below the top of the middle bands, $(\Delta/2 + w) -\Delta/2 = w$ when $w \ge \Delta$.

There are two different types of biphoton edge modes, \emph{conventional} and \emph{hybrid}. In a conventional edge mode, both photons excite the same single photon edge mode; thus $\beta = 0$. This mode is inevitably degenerate with modes belonging to the middle pair of biphoton Bloch bands, which are centred at $\beta = 0$. 

In a hybrid mode, one photon excites the edge mode, while the other excites a Bloch band mode; thus the hybrid modes form bands with $\beta = \pm |\mathbf{d}(k_s)|$ and $\beta = \pm |\mathbf{d}(k_i)|$. When the biphoton Bloch wave spectrum is gapped, hybrid modes with $w < |\mathbf{d}(k_i)| < \Delta$ reside in the gap.

We obtain the coupling efficiency by substituting the mode profiles Eq.~\eqref{general_modes} into Eq.~\eqref{coupling_efficiency},
\begin{align}
\Gamma_{p_s,p_i}(k_s,k_i) &= E^{(p)}_{1}(k_s + k_i ) \sin (\varphi_s / 2) \sin (\varphi_i / 2) e^{-i (\phi_s + \phi_i )} \nonumber \\
& + p_s p_i  E^{(p)}_{2}(k_s + k_i ) \cos (\varphi_s / 2) \cos (\varphi_i / 2), \label{2_band_efficiency}
\end{align}
with $p_{s,i} = \pm 1$ and $\varphi_{s,i}, \phi_{s,i} \equiv \varphi (k_{s,i}), \phi ( k_{s,i} )$. Recall $E^{(p)}_{1,2}(k)$ is the Fourier transform of the pump amplitude on the two different sublattices. $p_s p_i = 1(-1)$ if signal and idler photons come from the same (different) Bloch bands. This sign can always be absorbed into the relative phase of $E^{(p)}_{1,2}$, which means that, as far as the coupling efficiency is concerned, all the biphoton bands look the same, so the only difference will be in their dispersion (phase matching conditions).

Defining the vector
\begin{align}
\mathbf{n} &= \left( \sin (\varphi_s / 2) \sin (\varphi_i / 2) e^{-i (\phi_s + \phi_i )}, \right. \nonumber \\ & \qquad \quad \left. p_s p_i  E^{(p)}_{2}(k_s + k_i ) \cos (\varphi_s / 2) \cos (\varphi_i / 2) \right),
\end{align}
the angles $(\varphi_{s,i}, \phi_{s,i})$ define a direction on the Bloch sphere, and Eq.~\eqref{2_band_efficiency} can be recast as $\Gamma = \mathbf{E}^{(p)} (k_s + k_i ) \cdot \mathbf{n}$. Hence, the coupling efficiency is maximized when $\mathbf{E}^{(p)}(k_s + k_i)$ is parallel to $\mathbf{n}$, and zero if it is perpendicular. 

In a trivial phase $\varphi_{s,i}$ and $\phi_{s,i}$ do not exhibit any winding. If the gap is sufficiently large, they typically stay close to mean values independent of $k_{s,i}$, eg. Fig.~\ref{bloch_sphere}(b). Then it is possible to shape the pump profile $\mathbf{E}^{(p)}(k_s,k_i)$ such that all the modes in a band are strongly excited, resulting in behaviour similar to the homogeneous lattice case~\cite{SPDC_arrays}. On the other hand, in the nontrivial phase the winding of $\varphi_{s,i}$ or $\phi_{s,i}$ means that it is impossible to simultaneously excite all modes efficiently: shaping the pump profile such that $\mathbf{E}^{(p)}$ is parallel to $\mathbf{n}$ for some $k_{s,i}$, there are inevitably other values of $k_{s,i}$ for which they are perpendicular. Thus, there are selection rules preventing the excitation of some modes. This is the main consequence of nontrivial winding or topology in an infinite lattice.

So in summary, in 1D two band lattices the nontrivial topology has two main effects:
\begin{itemize}
\item when the single photon band gap is sufficiently large, there are hybrid biphoton edge modes with frequencies lying in the band gaps of the Bloch wave spectrum
\item the coupling efficiency for the excitation of Bloch waves is modulated, giving additional selection rules for the generation of biphotons
\end{itemize}

In the next Section we apply these ideas to a concrete example.

\section{Su-Schrieffer-Heeger model}
\label{sec:SSH_model}

\begin{figure}

\includegraphics[width=\columnwidth]{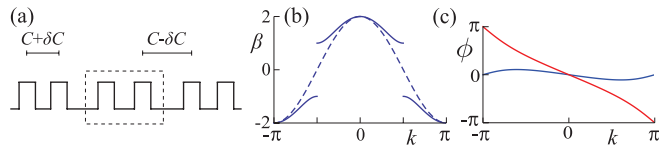}

\caption{(Color online) The 1D SSH model. (a) Lattice geometry with modulated waveguide separation (coupling), unit cell marked by dashed line. (b) Spectrum, shown in extended Brillouin zone scheme, for homogeneous ($\delta C = 0$, dashed) and dimerized ($|\delta C| = C/2$, solid) lattices. (c) Phase winding of eigenmodes in trivial ($\delta C = C/2$, blue) and nontrivial ($\delta C = -C/2$, red) phases.}

\label{SSH_model}

\end{figure}

The Su-Schrieffer-Heeger (SSH) model~\cite{SSH_paper,topological_review} presents a simple example of a 1D topological phase. Physically, it describes a 1D waveguide array where the waveguide separation is modulated such that the nearest neighbour coupling strength alternates between $C + \delta C$ and $C - \delta C$, see Fig.~\ref{SSH_model}(a). This is described by the single photon tight binding Hamiltonian
\begin{align}
\hat{H}^{(\mathrm{QW})} = \sum_n &\left( [C + (-1)^n \delta C ] \hat{a}_{n+1}^{\dagger} \hat{a}_n  \right. \nonumber \\
& \left. + [C - (-1)^n \delta C ] \hat{a}_{n-1}^{\dagger} \hat{a}_n \right),
\end{align}
Introducing two sublattices and applying a Fourier transform, we obtain the single photon Bloch Hamiltonian
\be 
\hat{H}^{(\mathrm{QW})}( k ) = [C + \delta C + (C - \delta C) \cos k ] \hat{\sigma}_x + (C - \delta C ) \sin k \hat{\sigma}_y, \label{SSH_Bloch_Hamiltonian}
\ee
which corresponds to Eq.~\eqref{2_band_hamiltonian} with $d_z = 0$. The spectrum,
\be 
\beta_{\pm}( k ) = \pm \sqrt{2} \sqrt{ C^2 + \delta C^2 + (C + \delta C)(C - \delta C ) \cos k }, \label{SSH_spectrum}
\ee
is plotted in Fig.~\ref{SSH_model}(b) in the extended Brillouin zone representation. When $\delta C = 0$ the model reduces to a homogeneous lattice with the usual single dispersion band. Nonzero $\delta C$ doubles the lattice period, forming two sublattices and splitting this band in two, each with a width of $2(|C| - |\delta C|)$, and separated by a gap of size $4|\delta C|$. 

Single photon Bloch functions are obtained as the eigenvectors of Eq.~\eqref{SSH_Bloch_Hamiltonian},
\begin{align} 
\mathbf{u}_{\pm} ( k ) &= \frac{1}{\sqrt{2}} \left( 1, \pm \frac{1}{\beta} [ (C - \delta C) e^{-ik} + (C + \delta C ) ] \right), \nonumber \\
&= \frac{1}{\sqrt{2}} \left( e^{-i \phi ( k ) }, \pm 1  \right),
\end{align}
so recalling Eq.~\eqref{general_modes}, $\varphi (k) = \pi / 2$, ie. Bloch functions live on the Bloch sphere's equatorial plane. The winding of $\phi( k )$, plotted in Fig.~\ref{SSH_model}(c), determines the lattice topology. In the trivial phase $\delta C > 0$, intracell coupling is stronger and $\phi ( k )$ shows no winding, with $\phi(k) \approx 0$ for all $k$. In the nontrivial phase $\delta C < 0$, intercell coupling is stronger and $\phi (k ) - \phi (k + 2\pi ) = 2 \pi$, winding once around the Bloch sphere's equatorial plane. Note that in the limit of an infinite lattice, the choice of unit cell boundary is arbitrary, so this ``topology'' becomes ill-defined. However, in any \emph{finite} lattice, the nontrivial phase is distinguished by a pair of ``topologically protected'' edge modes at $\beta = 0$ (the middle of the band gap)~\cite{topological_review}. These modes appear when a strong bond is broken to form the edge. Experimentally, one can compare the two phases using a single lattice with the two edges terminated differently.

\begin{figure}

\includegraphics[width=\columnwidth]{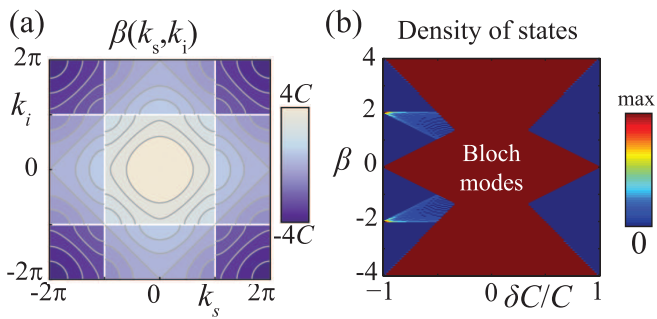}

\caption{(Color online) (a) Biphoton Bloch wave spectrum, shown in extended Brillouin zone scheme, $\delta C / C = 1/2$. (b) Density of states as a function of coupling modulation $\delta C / C$. Bloch bands (solid red region) are gapped for $|\delta C / C | > 1/3$. Edge modes only exist in the nontrivial phase, $\delta C / C < 0$.}

\label{biphoton_bands}

\end{figure}

Biphoton modes $\beta_{n_s,n_i} (k_s,k_i)$ are constructed as combinations of pairs of single photon modes. The biphoton spectrum is shown in Fig.~\ref{biphoton_bands}(a), once again using the extended Brillouin zone representation. In contrast to the single photon case, here the spectrum remains gapless for $|\delta C / C | < 1/3$ (as long as the width of a single photon band exceeds the band gap).

The nontrivial phase hosts pairs of biphoton bands at each edge corresponding to the ``hybrid'' edge modes, shown in Fig.~\ref{biphoton_bands}(b). These bands bifurcate from the the topological phase transition at $\delta C = 0$. However, for $-1/3 < \delta C / C < 0$ they still overlap with the bulk Bloch bands. Notice how their edges remain pinned at $\beta = \pm 2$. This is because the ``topological protection'' ensures the single photon end mode remains fixed at $\beta = 0$. In contrast, the trivial phase $\delta C > 0$ does not have any edge modes

\subsection{Pumping - infinite lattice}

Here we consider in-phase pumping of two adjacent waveguides far from the lattice edges (the infinite lattice limit). We will start by considering examples of numerical solutions of Eq.~\eqref{schrodinger_solution} before discussing how the results generalise. In the following examples, we use $|\delta C / C | = 1/2$, and assume a normalised propagation distance $L = 10/C$.

\begin{figure}

\includegraphics[width=\columnwidth]{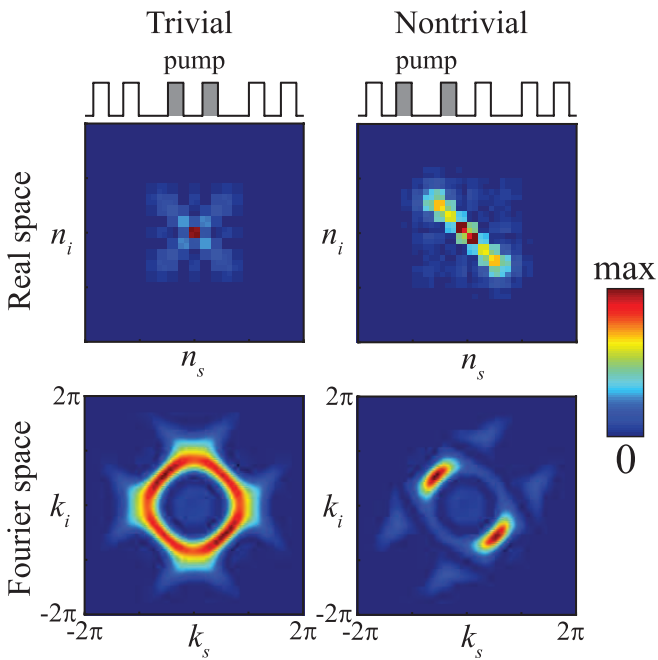}

\caption{(Color online) Real and Fourier space biphoton correlations, trivial (left) and nontrivial (right) phases. $|\delta C|=C/2$. $LC = 10$. $\Delta \beta^{(0)} = 3 C$ (band 1 excitation).}

\label{inphase_corr}

\end{figure}

Fig.~\ref{inphase_corr} shows the output biphoton correlations in the trivial and nontrivial phases when the pump frequency is resonant with the first biphoton band, $\Delta \beta^{(0)} = 3C$. The output in the trivial phase displays bunching and antibunching, resembling the output of a homogeneous lattice with a single waveguide pump~\cite{SPDC_arrays}. In contrast, pronounced antibunching occurs in the nontrivial phase; photon bunching is strongly suppressed. 

This result can be understood quite intuitively by considering the strong modulation limit $|\delta C| \approx C$, in which the lattice consists of strongly coupled ``dimer'' pairs of waveguides, with weak coupling between neighbouring dimers. At each dimer, the single waveguide modes hybridize to form in- and out-of-phase modes. When $\delta C > 0$, the pump excites the in-phase mode of a single dimer; thus the output resembles that of a homogeneous lattice when a single waveguide is excited. In the nontrivial phase, the pump excites two dimers; interference between them suppresses photon bunching.

\begin{figure}

\includegraphics[width=\columnwidth]{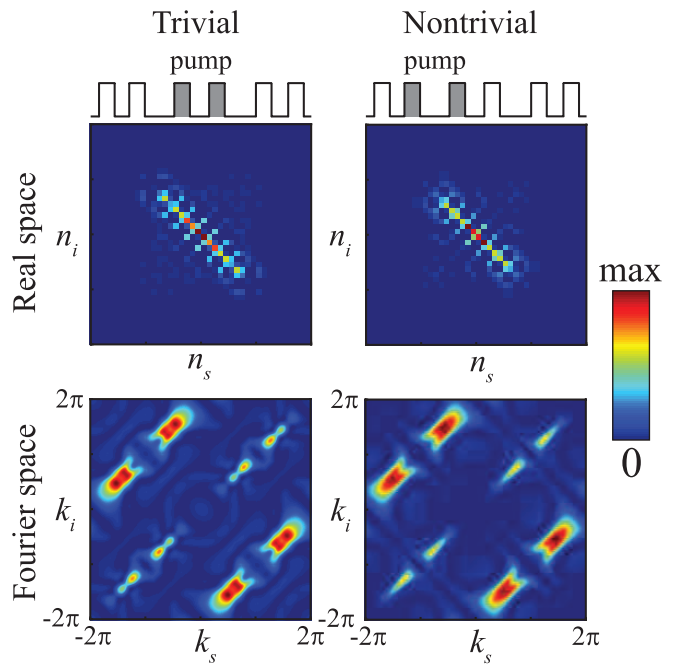}

\caption{(Color online) Real and Fourier space correlations, trivial (left) and nontrivial (right) phases. $|\delta C|=C/2$. $LC = 10$. $\Delta \beta^{(0)} = 0$ (bands 2 \&3 excitation).}

\label{hybrid_corr}

\end{figure}

The response changes when the pump is tuned to the 2nd and 3rd (overlapping) bands. The correlations in Fig.~\ref{hybrid_corr} no longer show any significant qualitative difference between the trivial and nontrivial phases; both display strong antibunching. The main quantitative difference is that the total output intensity is an order of magnitude smaller in the trivial phase.

Let us now relate these observations back to the general theory. The pump detuning $\Delta \beta^{(0)}$ imposes a phase-matching condition on the spatial modes, such that only Bloch modes in resonance can be strongly excited. However, the Bloch mode spectra for $\delta C$ and $-\delta C$ are identical, so the coupling efficiency $\Gamma$ is solely responsible for the differences in the biphoton correlations. Evaluating Eq.~\eqref{2_band_efficiency}, we obtain
\be 
\Gamma = \frac{1}{2} \left( E_1^{(p)} (k_s + k_i) e^{-i [ \phi (k_s) + \phi ( k_i) ]} + p_s p_i E_2^{(p)} (k_s + k_i)  \right), \label{SSH_efficiency}
\ee
where $p_{s,i} = \pm$, and $E_{1,2}^{(p)}(k_s+k_i)$ is the Fourier transform of the pump amplitudes on the two sublattices. Since only a single unit cell is pumped, the Fourier transform is a constant, ie. $E_{1,2}^{(p)}(k_s+k_i) = E_{1,2}^{(p)}$, independent of $k_{s,i}$. We see that the topology, via the phase $\phi(k_{s,i})$, affects the interference between the two sublattices, which in turn controls $\Gamma$. When a single sublattice is pumped (eg. $E_{2}^{(p)} = 0$), no interference occurs, so $|\Gamma|$ is independent of $\phi$ and $\delta C$ and the topology is irrelevant. So both sublattices must be pumped to observe any sensitivity to the topology.

In the nontrivial phase, as either $k_s$ or $k_i$ traverse the Brillouin zone, $\phi(k_{s,i})$ takes \emph{all} possible values. Hence there is \emph{always} a curve through the $(k_s,k_i)$ Brillouin zone where $|\Gamma|$ attains its maximum of $(|E_{1}^{(p)}| + |E_{2}^{(p)}|)/2$ (constructive interference), and one where $|\Gamma|$ drops to its minimum of $(|E_{1}^{(p)}| - |E_{2}^{(p)}|)/2$ (destructive interference). When $|E_{1}^{(p)}| = |E_{2}^{(p)}|$ this minimum is zero, so we have perfect destructive interference and coupling into the corresponding Bloch wave is forbidden. Changing the relative phases of $E_{1,2}^{(p)}$ shifts the two curves, but it does not remove them. In contrast, in the trivial phase $\phi(k_{s,i})$ does not display any winding, $\phi(k_{s,i}) \approx 0$, and the efficiency is only controlled through the relative phase $\theta$ of $E_{1,2}^{(p)}$. 

\begin{figure}

\includegraphics[width=\columnwidth]{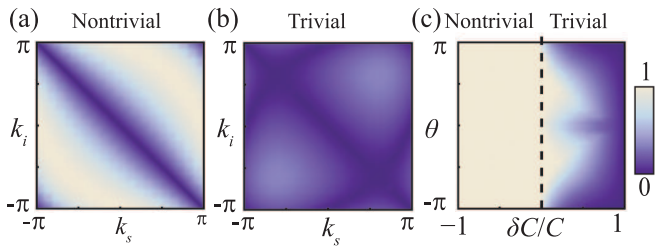}
\caption{(Color online) SPDC coupling efficiency $|\Gamma (k_s,k_i)|$ for two waveguide pumping, normalised by pump intensity, equal pump intensities, $\pi$ relative phase between the two waveguides, in nontrivial (a) and trivial (b) phases. $|\delta C / C | = 1/2$. The nontrivial phase displays a maximum, and minimum of zero along curves in 2D Brillouin zone. In trivial phase, these lines are not present, $|\Gamma|$ is controlled mainly by the relative pump phases. (c) Contrast in the coupling efficiency, max($|\Gamma (k_s,k_i)|$) - min($|\Gamma (k_s,k_i)|$), as a function of relative pump phase $\theta$ and coupling modulation $\delta C$}

\label{gamma_plots}

\end{figure}

We demonstrate these two different cases by plotting $|\Gamma (k_s,k_i)|$ in Fig.~\ref{gamma_plots}(a,b), assuming out-of-phase pumping of the two waveguides and phase matching with band 1. In this case, $|\Gamma|$ vanishes for both phases when $k_i = -k_s$, ie. antibunching is suppressed. In the trivial phase however, $|\Gamma$| remains small for all $(k_s,k_i)$, ie. no modes are efficiently excited, while $|\Gamma|$ attains its maximum of 1 in the nontrivial phase and efficiently excites some modes. This explains the results in Fig.~\ref{hybrid_corr} (similar correlations, but different intensities).

More generally, we show in Fig.~\ref{gamma_plots}(c) the effect of $\delta C$ and $\theta$ by plotting the contrast, max($|\Gamma (k_s,k_i)|$) - min($|\Gamma (k_s,k_i)|$), for the case $|E_{1}^{(p)}| = |E_{2}^{(p)}|$. We verify the reasoning in the previous section that in the nontrivial phase, the contrast is always maximum, while in the trivial phase it decreases to zero. 

\begin{figure}

\includegraphics[width=\columnwidth]{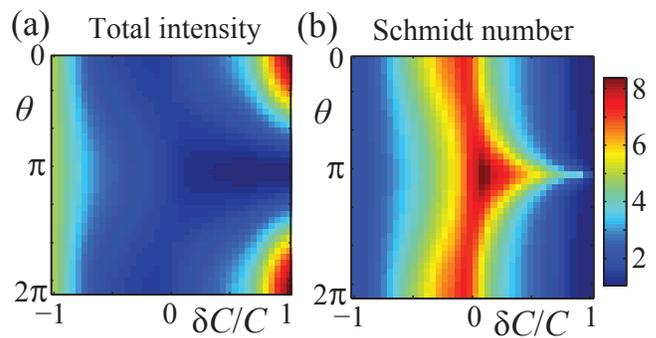}
\caption{(Color online) (a) Total intensity of down-converted photons and (b) Schmidt number as a function of $\delta C / C$ and relative phase $\theta$ of the two pumped waveguides. $\Delta \beta^{(0)} = 2(|C| + |\delta C|)$ (middle of band 1 excitation).}

\label{scans}

\end{figure}

The total contrast controls the efficiency of the SPDC. Fig.~\ref{scans}(a) shows the total down-converted intensity as a function of $\delta C / C$ and $\theta$. In the nontrivial phase, there are always some modes that are spatially matched with the pump and thus strongly excited. Hence the total output intensity is relatively insensitive to the relative pump phase $\theta$. In the trivial phase, it is crucial that $\theta$ matches the selected band's Bloch wave profile, otherwise no spatial modes are strongly excited and the down-converted intensity vanishes. 

Fig.~\ref{scans}(b) shows the influence of $\theta$ and $\delta C / C$ on the Schmidt number of the biphoton state, assuming the pump frequency is tuned to the centre of band 1. The lattice topology has a less significant effect here; in fact the Schmidt number is largest when $\delta C \approx 0$ (a homogeneous lattice). 

So far we have focused on a pump that is confined to a single unit cell of the lattice (two waveguides). Let us now consider briefly the effect of a broader pump beam. As the pump profile is made wider, it becomes more localized within the lattice's Brillouin zone, such that $|E_{1,2}^{(p)}| = |E_{1,2}^{(p)}|(k_s + k_i)$ acquires a $k$-dependence.

Applying Eq.~\eqref{SSH_efficiency}, this localization of the pump around some point in the Brillouin zone leads to an additional, $k_{s,i}$-dependent modulation of $|\Gamma|$. Notice however that this modulation is distinct from that arising from the Bloch waves themselves: it depends on the sum of the down-converted photon wavenumbers, $k_s + k_i$, instead of $k_{s,i}$ individually. Furthermore, this additional modulation induced by a broad beam is independent of the lattice properties, including its topology.

\begin{figure}

\includegraphics[width=\columnwidth]{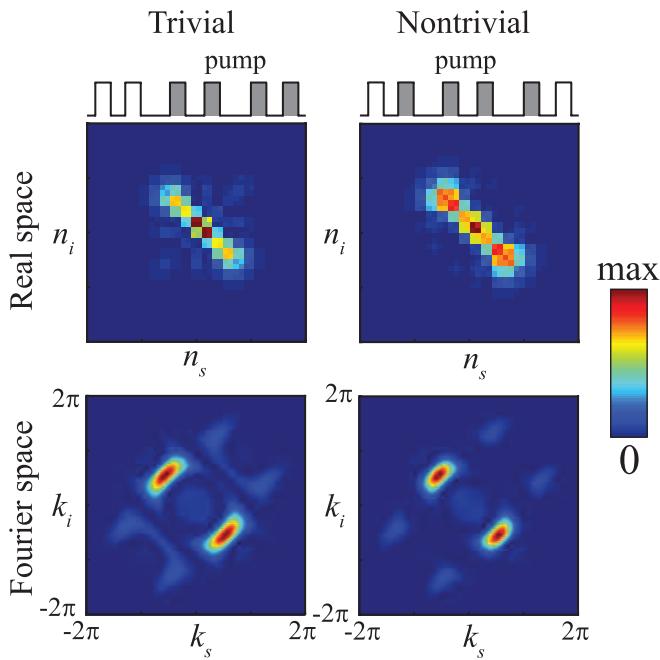}

\caption{(Color online) Real and Fourier space correlations, trivial (left) and nontrivial (right) phases. $|\delta C|=C/2$. $LC = 10$. $\Delta \beta^{(0)} = 3C$ (band 1 excitation). Four waveguides pumped.}

\label{wide_corr}

\end{figure}

\begin{figure}

\includegraphics[width=\columnwidth]{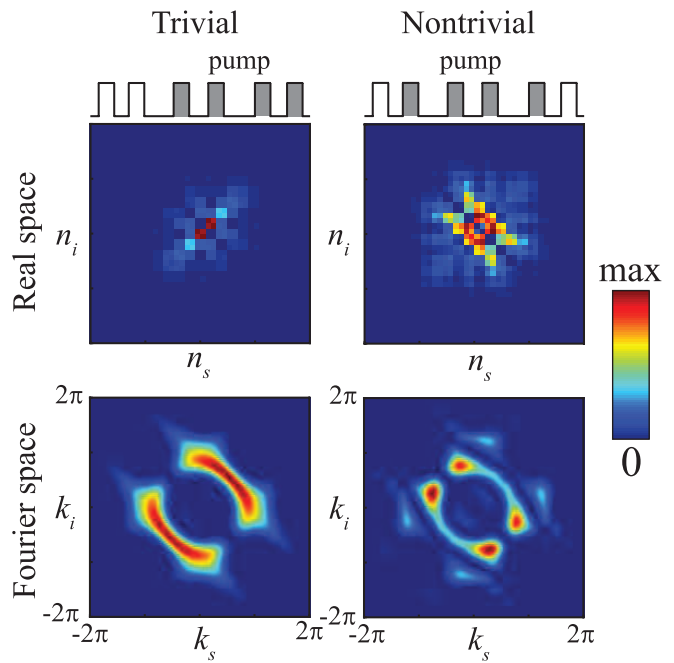}

\caption{(Color online) Real and Fourier space correlations, trivial (left) and nontrivial (right) phases. $|\delta C|=C/2$. $LC = 10$. $\Delta \beta^{(0)} = 0$ (bands 2\&3 excitation). Four waveguides pumped, $\pi$ phase difference between unit cells.}

\label{wide_corr_2}

\end{figure}

As an example, we consider two examples of biphoton correlations when two unit cells (four waveguides) are pumped. Pumping the waveguides in phase in Fig.~\ref{wide_corr} favours antibunching~\cite{SPDC_arrays}. The selection rule imposed by the lattice topology (which favours antibunching only in the nontrivial phase) becomes redundant, and both phases display similar correlations and antibunching. Conversely, pumping the two unit cells with a $\pi$ phase difference in Fig.~\ref{wide_corr_2} promotes photon bunching~\cite{SPDC_arrays}, clearly visible in the trivial phase. In the nontrivial phase, this additional selection rule competes with topology-imposed suppression of bunching to produce a complex pattern of correlations.

\subsection{Pumping - edge of lattice}

\begin{figure}

\includegraphics[width=\columnwidth]{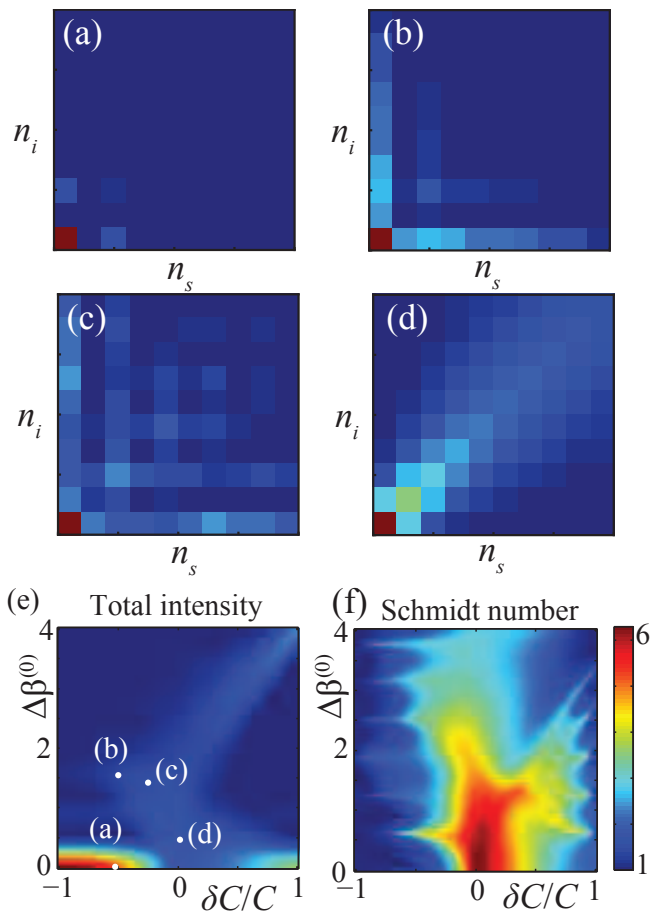}

\caption{(Color online) (a-d) Real space biphoton correlations when edge of lattice is pumped. Parameters indicated in (e). (a) Pump resonant with conventional edge mode. (b) Pump resonant with hybrid edge mode. (c) Pump resonant with hybrid edge mode and Bloch waves. (d) Homogeneous lattice with no edge modes. (e) Total biphoton intensity and (f) Schmidt number of the biphoton state as a function of pump detuning and coupling modulation.}

\label{edge_pump}

\end{figure}

We next consider a finite lattice and the role of the topologically protected edge modes. Fig.~\ref{edge_pump}(a-d) shows the real space output biphoton correlations when the waveguide at the end of the lattice is pumped, for different modulation strengths $\delta C$ and pump detunings $\Delta \beta^{(0)}$.

First, when the pump is tuned to the conventional edge mode, $\Delta \beta^{(0)} = 0$, we observe strong localization of the output in Fig.~\ref{edge_pump}(a), even though the pump is also resonant with Bloch waves. This is because of the strong overlap of the pump profile with the edge mode.

If the pump is tuned to the centre of the hybrid edge mode band, $\Delta \beta^{(0)} = \sqrt{2 (C^2 + \delta C^2)}$, there are four distinct regimes depending on $\delta C$:
\begin{itemize}
\item $\delta C / C > 1/3$ (trivial and gapped): The pump is tuned to a band gap, so no modes are resonantly excited.
\item $0 < \delta C / C < 1/3$, (trivial and gapless). The pump resonantly excites bulk biphoton modes, which propagate away from the edge. 
\item $-1/3 < \delta C / C <0$, (nontrivial and gapless). The pump resonantly excites bulk modes and an edge mode. 
\item $\delta C < -1/3$, (nontrivial and gapped). Only a biphoton edge mode is resonantly excited. One photon is trapped at the edge, while the other propagates into the bulk.
\end{itemize}
Fig.~\ref{edge_pump}(b,c) demonstrates the last two regimes.

For comparison Fig.~\ref{edge_pump}(d) shows the output correlations when $\delta C = 0$ (homogeneous lattice) and there are no edge modes. Photon bunching occurs, with signal and idler both propagating into the bulk.

We consider more generally in Fig.~\ref{edge_pump}(e,f) how the biphoton intensity and Schmidt number depend on the pump detuning and coupling modulation.  Due to the strong overlap with the pump beam, the output intensity is maximum when the pump is resonant with the conventional edge mode. However, since only a single mode is strongly excited, the Schmidt number $K \approx 1$ reveals there is no entanglement. Similar to the bulk case, $K$ is maximized for relatively small $\delta C$, when the pump is tuned to the centre of the Bloch bands.

In summary, we have shown how the SSH model can exhibit nontrivial topology in biphoton correlations: in the bulk (additional selection rules), and at the edge (``hybrid'' biphoton edge modes).

\section{Modulated lattice depth}
\label{sec:modulated_array}

\begin{figure}

\includegraphics[width=\columnwidth]{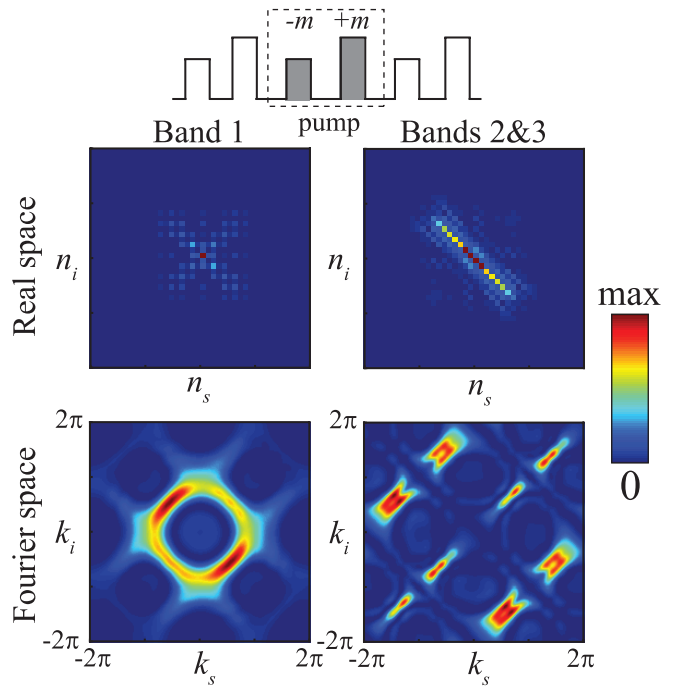}

\caption{(Color online) Biphoton correlations in a binary lattice, tuned to have same Bloch wave spectrum as the SSH model with $\delta C / C = 1/2$. (a,b) $\Delta \beta^{(0)} = 3 C$, (c,d) $\Delta \beta^{(0)} = 0$. Compare against Fig.~\ref{inphase_corr} and Fig.~\ref{hybrid_corr} respectively and observe the resemblance to the trivial phase.}

\label{massive_correlations}

\end{figure}

For comparison, we briefly consider here an experimentally accessible ``nontopological'' model: a binary lattice where the waveguide depths are modulated with strength $m$, while the waveguide spacing and coupling strength $C^{\prime}$ are constant, see Fig.~\ref{massive_correlations}.

Following the same procedure as for the SSH model, we obtain the Bloch Hamiltonian
\be 
\hat{H}^{(\mathrm{QW})}(k) = C^{\prime} (1 + \cos k) \hat{\sigma}_x + C^{\prime} \sin k \hat{\sigma_y} + m \sigma_z,
\ee
with spectrum
\be 
\beta_{\pm} = \pm \sqrt{ 2 C^{\prime 2} ( 1 + \cos k ) + m^2  }, \label{eq:onsite_spectrum}
\ee
and Bloch functions
\begin{align}
\mathbf{u}_{\pm} ( k )&= \frac{1}{\sqrt{2} \mathcal{N}} \left( m \pm \beta, C^{\prime} (1 + e^{i k} ) \right), \label{onsite_eigenvectors} \\
\mathcal{N}^2(k) &= m ( m \pm \beta ) + 2 C^{\prime 2} ( 1 + \cos k ).
\end{align}
We set $m = 2 \delta C$ and $C^{\prime} = \sqrt{C^2 - \delta C^2}$, so that the bulk spectrum is identical to the SSH model's Eq.~\eqref{SSH_spectrum}, allowing for a fair comparison. The only remaining difference between the two models is the singular winding of the Bloch functions in the SSH model, which is absent here.

At $k = 0$, the Bloch functions excite both sublattices, while at $k = \pi$ (Brillouin zone edge), they reside on a single sublattice only, with energies $\pm m$. Fig.~\ref{bloch_sphere}(a) shows the Bloch functions using the Bloch sphere representation. Since the curve can be continuously shrunk to a point, there is no nontrivial winding as the Brillouin zone is traversed. This is because taking the limit $m \rightarrow \infty$ continuously deforms the lattice to an effectively homogeneous chain without closing the band gap: one can see this by Taylor expanding the dispersion relation Eq.~\eqref{eq:onsite_spectrum} for small $C^{\prime} / m$ and recovering a $\beta \sim \cos k$ dispersion relation, with the Bloch functions independent of $k$. In a finite lattice, there are no edge modes.

We consider similar to Figs.~\ref{inphase_corr},\ref{hybrid_corr} pumping two adjacent waveguides far from the lattice edge. Similar to the trivial phase in the SSH models, the output correlations in Fig.~\ref{massive_correlations} resemble those from pumping a single waveguide in a homogeneous lattice: pumping the 1st band reveals both bunching and antibunching, while antibunching is favoured in bands 2 and 3. 

This example further highlights how lattices can produce very different biphoton correlations even when their spectra (eigenvalues) are identical and they are pumped in the same way, because of the additional selection rules imposed by the Bloch functions and their trivial or nontrivial topology.

\section{Conclusions and outlook}
\label{sec:conclusion}

In summary, we have explored the effect of lattice topology on spontaneous parametric down-conversion in one dimensional quadratic nonlinear waveguide arrays. We have shown how nontrivial winding in the Bloch wave spectrum leads to selection rules for the generation of entangled photon pairs. Finite lattices can host topologically protected edge modes, which interestingly enable the generation of entanglement between localized and propagating spatial modes. As a specific example we considered in detail an analogue of the Su-Schrieffer-Heeger model, which can be experimentally realised in lithium niobate nonlinear waveguide arrays using existing fabrication techniques.

The study of extensions to two dimensional topological phases remains an open problem. Can the biphoton spectrum $\beta ( k_s,k_i)$ and its eigenmodes host genuinely two dimensional effects, such as nonzero Chern number? Presumably, the ``edge modes'' in such a system would involve one photon bound at the edge, with the other propagating into the bulk. Another possible avenue to explore is SPDC in two dimensional waveguide arrays with nonzero Chern number. A two dimensional array results in a four dimensional biphoton spectrum, which raises the intriguing possibility of emulating highly exotic topological phases, such as the four dimensional quantum Hall effect~\cite{4d_hall}. However, experimental realisations would be quite challenging, since so far SPDC has been limited to one dimensional nonlinear waveguide arrays.

\section*{Acknowledgements}

This work has been supported by the Australian Research Council, including Discovery Project No. DP130100135 and Future Fellowship No. FT100100160.

\end{document}